# Kinetic and Thermodynamic Limits of Sodium Pore Filling in Hard Carbon

Ihor Radchenko[2], Andrés Felipe Usuga[2], Yuxin Tang[3*], Oleksandr I. Malyi[1*]

[1]Qingyuan Innovation Laboratory, Quanzhou, P. R. China

[2]Centre of Excellence ENSEMBLE3 Sp. z o. o., Warsaw, Poland

[3]College of Chemical Engineering, Fuzhou University, Fuzhou, China

*Corresponding author; email: yxtang@fzu.edu.cn (Y. Tang), oleksandrmalyi@gmail.com (O.I. Malyi)

**Abstract:** The low-voltage plateau-like capacity of hard carbon anodes is widely regarded as a key signature of sodium storage in closed pores, often defining the performance of Na-ion batteries. Here, we combine first-principles calculations with a grand canonical Monte Carlo pore-filling model to show that Na storage in closed pores is not governed by a single equilibrium-plateau mechanism but by a distribution of pore-specific thermodynamic and kinetic activation processes. Thermodynamically, the voltage at which Na pore filling becomes favorable decreases with increasing pore size. Kinetically, pore filling involves Na nucleation, coalescence, growth, and bulk diffusion. While small pores fill rapidly, large pores exhibit both a smaller thermodynamic stimulus and significant kinetic limitations. Nucleation can be accelerated by introducing defects at the pore walls, although an excessively high defect concentration slows bulk diffusion. This work shows that not all closed pores contribute equally to Na storage: some may be thermodynamically favorable but kinetically inactive under realistic charge/discharge rates. Because of this, the inverse design of hard carbon electrodes should target a kinetically accessible pore-size and defect-density window that balances high plateau-like capacity with fast-charging performance.

## 1. Introduction

Hard carbon has emerged as one of the most promising anode materials for Na-ion batteries due to its low cost, environmental friendliness, and excellent cyclability.[1,2] Unlike graphite, which solely relies on an intercalation storage mechanism (e.g., Li insertion between graphite layers), the charge storage within the hard carbon electrode is defined by a corresponding contribution of different local structural motifs present in a complex structure of misaligned graphite-like domains enriched by open and closed pores of different sizes and degrees of accessibility.[3,4] For instance, for Na storage, the voltage charge-discharge profile exhibits slope- and plateau-like regions that correspond to multiple Na storage mechanisms (Figure 1a,b). Traditionally, it is believed that plateau capacity directly correlates with the size[5,6] and volume[6,7] of pores (specifically, closed pores), suggesting that pore filling is the dominant capacity-limiting mechanism in this low-voltage plateau region. Moreover, the pore size may also influence the shape of the voltage profile as different pore sizes have been shown to be filled at different states of charge in hard carbon.[5] In contrast, Na sorption on defects and Na intercalation between expanded graphite layers are expected to mainly contribute at high voltages,[8] although more recent investigations reveal that defects can additionally play a role at lower voltages.[9] Such understanding of Na storage mechanisms led to the development of different strategies (e.g., lattice reconstruction strategy,[4] data-driven intelligent carbonization,[10] rapid high-temperature heating,[11] introducing pore-forming agents[12]) to optimize the plateau storage capacity or fast-charging performance. However, often, such strategies are based on tuning pyrolysis and carbonization stages to optimize the electrochemical performance without a deep understanding of the role of each local structural motif in Na storage or how tuning specific synthesis conditions modifies the distribution of such motifs. Indeed, from a simple thermodynamic perspective, maximizing the plateau Na-storage capacity would require increasing the size of the closed pores and expanding the interlayer distance while minimizing the concentration of defects that contribute to Na storage at high voltages. Although this strategy offers a route towards maximizing the theoretical limit for Na-storage capacity, under typical battery operating conditions, kinetic factors also play a significant role, often preventing this theoretical limit from being reached.[4] Plateau-type hard carbon electrodes can deliver substantial low-voltage capacity at low charge/discharge rates, but this contribution often decreases during fast charging. This potentially indicates that not all pores are kinetically activated for Na storage upon fast charging.[5] This distinction is critical because the conventional goal of maximizing closed-pore volume implicitly assumes that every closed pore contributes equally to reversible Na storage. In practice, however, while Na filling of a given pore can be thermodynamically favorable, the pore can still remain inactive for Na storage under realistic cycling conditions if pore-filling kinetics are too slow. Therefore, from a fundamental perspective, the key design parameter is not simply the total closed-pore volume or the largest achievable pore size, but the fraction of pores that can be filled within the electrochemical timescale of battery operation. Motivated by this observation, herein, we employ first-principles calculations and Monte Carlo simulations to investigate the thermodynamic driving forces and kinetic limitations governing Na pore filling (see Supplementary Information and references therein[8,13–35] for all the technical details). Specifically, based on the obtained results, we demonstrate that hard carbon containing a range of different-sized pores does not exhibit a plateau region in the conventional sense (with a constant voltage). Instead, at low voltages, the voltage profile is rather weakly sloped, where each point corresponds to the activation of a different pore size (i.e., smaller pores are activated at higher voltages, as illustrated in Figure 1b). In this light, the experimentally observed plateau corresponds to an averaged feature arising from filling of different closed pores rather than a truly flat potential associated with a single storage mechanism. Importantly, we define critical stages in pore filling, including Na nucleation, coalescence, growth, and diffusion-controlled filling propagation (Figure S1), demonstrating that each of these processes is highly sensitive to the pore size (Figure 1c). Our results further reveal that large pores may remain kinetically inaccessible owing to intrinsically limited diffusion and low thermodynamic driving force (undervoltage), thereby defining a critical pore size above which pores can no longer actively participate in Na storage under practical operating conditions. Defects at the pore walls may widen the kinetically accessible pore size window, but the exact defect density needs to be optimized to allow larger pores to be filled. Thus, by showing that closed pores should not simply be made larger to maximize Na storage in the thermodynamic limit but should be engineered within a kinetically accessible size window to balance plateau-like capacity and fast-charging performance, this work provides a foundation for inverse design of next-generation hard-carbon electrodes.

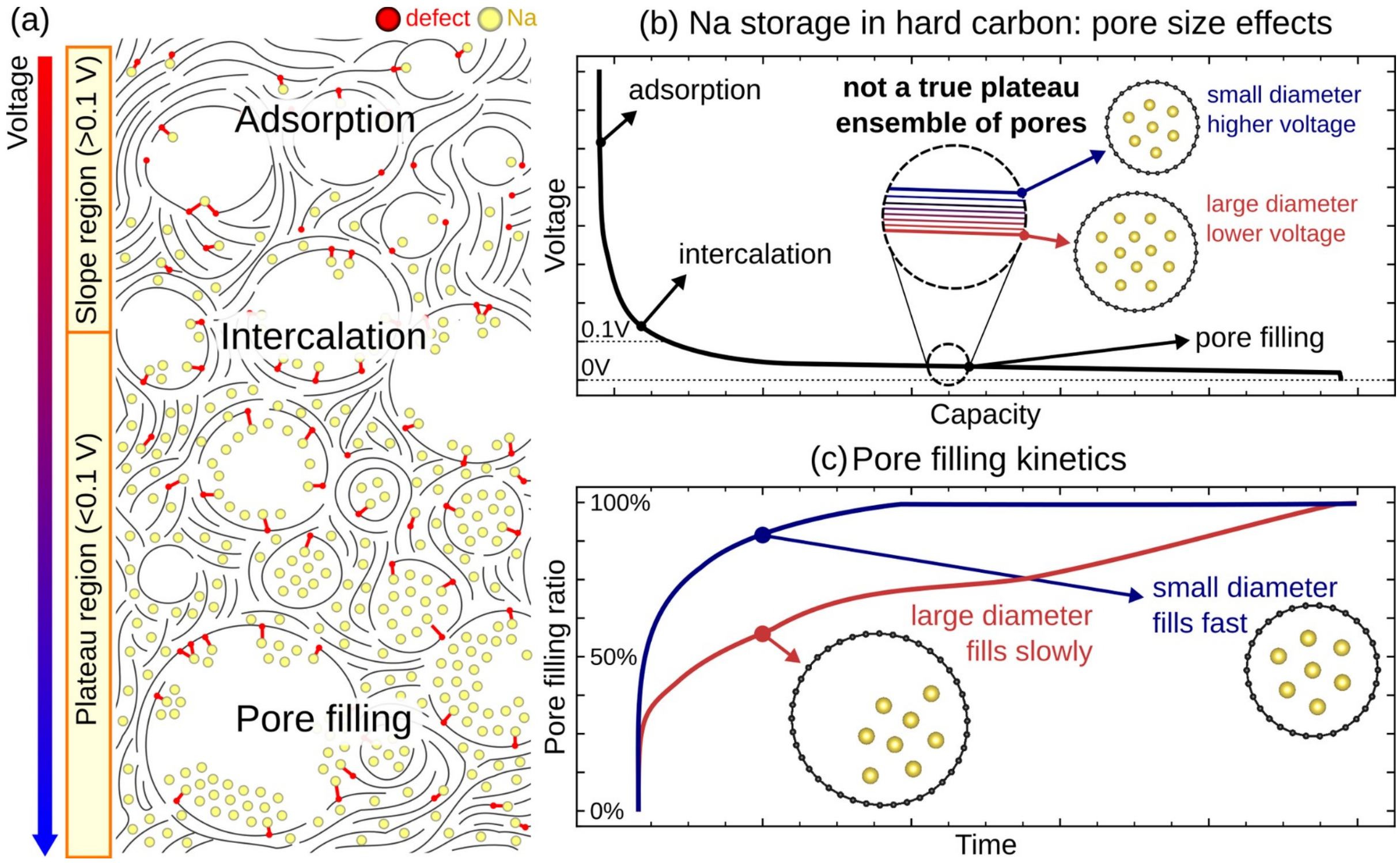

**Figure 1.** Na storage mechanism in hard carbon, illustrating pore filling as a critical part of the plateau capacity. (a) Schematic illustration of hard carbon structure showing adsorption, intercalation, and pore-filling contributions for Na storage in a hard carbon electrode. (b) Schematic illustration of the voltage profile for hard carbon, presenting the main Na storage mechanisms and the joint contribution of different pore sizes to the voltage plateau. (c) Small and large pores exhibit different pore-filling kinetics, as shown later by Monte Carlo simulations.

## 2. Results and discussion

Multiple experimental studies indicate that hard carbon electrode materials exhibit plateau capacity, largely determined by Na storage in closed pores.[5–7] In the thermodynamic limit, this implies that the ion chemical potential remains constant over a range of compositions,[36] which is classically associated with two-phase coexistence, like in the case of $Li_4Ti_5O_{12}$ and $Li_7Ti_5O_{12}$ for Li-ion batteries.[37] We note, however, that despite common discussion in the literature of hard carbon electrode materials, a single, well-defined two-phase transition governing the entire plateau region is unlikely, as hard carbon represents a structurally heterogeneous material where multiple local motifs coexist (Figure 1a). Indeed, in contrast to an ideal thermodynamic plateau, the low-voltage region of hard-carbon anodes is more accurately described as a plateau-like, weakly sloped process. For instance, Stratford et al.[38] showed that hard carbons exhibit a sloping high-voltage process followed by a "plateau-like" lower-voltage process, whose onset is sample-dependent and can be operationally defined from differential-capacity analysis rather than by the commonly employed fixed 0.1 V cutoff. Indeed, given the distribution of local motifs and continuous distribution of pore sizes,[5,39] it is unclear how a single, well-defined two-phase reaction is even possible in this context. Motivated by this, herein we revisit the Na storage mechanism in the plateau region using the simplest pore model. Specifically, we consider a set of defect-free carbon nanotubes with different radii using density-functional theory (DFT), random-structure sampling, and a genetic algorithm (details in Supplementary Information and Figure S2 and S3 therein) to identify the lowest-energy arrangement of Na atoms at different concentrations within the pore (we note that while this model is a simplified approximation for the system, it provides a key physical picture that can be directly used for interpretation of the storage mechanism). The obtained results are shown in Figure 2 and demonstrate a few common tendencies: (i) a pore, even without defects, can store Na (this is different from recent work that reports DFT calculations for systems without accounting for van der Waals (vdW) interaction,[20] see Figure S2 in the Supporting Information highlighting the role of vdW correction); (ii) favorable states appear at finite Na concentration, suggesting that Na atoms become stable as clustered/pore-filled arrangements and not as independent single Na atoms interacting with C atoms; (iii) smaller pores stabilize low Na concentration at higher voltages, whereas larger pores can accommodate higher Na concentration but contribute at lower voltages. This trend indicates a clear confinement effect: in small pores, the interaction between Na and the carbon framework is sufficiently strong to stabilize the clustered state, whereas in larger pores the stabilization becomes too weak to compensate for the reference energy of bulk Na. In terms of battery operation, these results directly imply that different pore sizes contribute differently to the battery voltage; specifically, large pores contribute to lower voltages than smaller ones. Considering the fact that hard carbon contains a distribution of local pores,[5,39] the low-voltage plateau in hard carbon is more plausibly interpreted as an ensemble-averaged response of a structurally heterogeneous carbon network, in which pores of different sizes become favorable for Na storage over a narrow but finite voltage window. Moreover, small pores can be thermodynamically activated at voltages higher than 0.1 V (i.e., the voltage typically associated with the onset of pore filling). Moreover, because a finite Na concentration is needed to achieve a stable configuration (Figure 2b), there is a nucleation-like threshold for pore filling. Isolated or very small Na populations are not sufficiently stabilized, whereas clustered Na configurations become favorable. Larger pores may require more Na atoms to reach this stable regime, consistent with slower filling kinetics, as will be shown later.

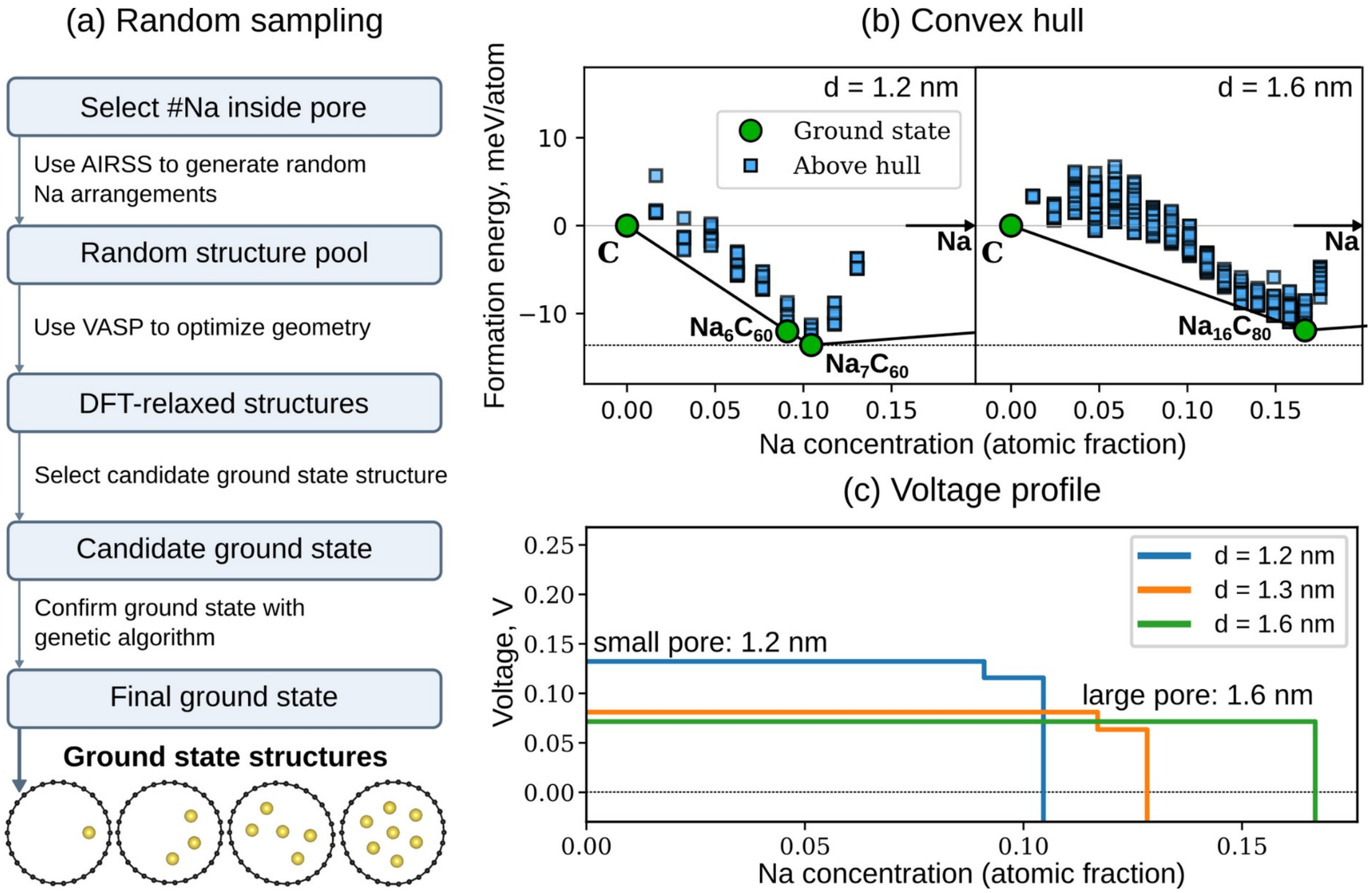

**Figure 2.** Pore-size dependence of sodium filling energetics and voltage. (a) Workflow used to search for optimal Na-inside-pore configurations. The optimal configurations are found using Ab initio Random Structure Searching (AIRSS) [26,27], further confirmed by a genetic algorithm [28]. (b) Sample convex hulls for the pores with 1.2 nm and 1.6 nm diameters obtained by combining all the structures considered in (a); see more details in Figure S2. (c) Na voltage profiles for different pore sizes. For all calculations, the pore is modeled as a CNT of different radii (i.e., CNT(15,0), CNT(17,0), and CNT(20,0) for pores with diameters of 1.2, 1.3, and 1.6 nm, respectively).

In the traditional sense of Na storage, defects are usually associated with high-voltage storage, and plateau capacity is believed to be dominated mainly by pores and, partially, by Na intercalation into crystalline domains.[7,8,40] Indeed, defect contribution to high voltages is evident from our Na binding energy calculations for the defects, where, for instance, the Na adsorption energy (relative to metallic Na) on a graphene monovacancy is -1.01 eV (contributing to a voltage of 1.01 V). In this light, it is widely believed that defect concentration needs to be reduced to avoid battery slope regions and maximize initial Coulombic efficiency.[3,41,42] However, defects can also play a role in other Na storage mechanisms in hard carbon. Defects and Na-passivated defects can act as centers for nucleation and growth of metal clusters. According to our ab initio molecular dynamics (AIMD) calculations (Figure 3), randomly placed Na atoms on defective graphene diffuse and tend to form metallic clusters. The corresponding clustering energy (energy of a Na cluster relative to a set of isolated Na atoms on graphene away from defects) also indicates that Na clustering on defects is favorable: the computed clustering energies per Na atom for Na, $Na_3$, $Na_5$, $Na_8$, and $Na_{12}$ clusters are -1.28, -0.62, -0.48, -0.34, and -0.30 eV per Na atom, respectively (see inset in Figure 3e). Even the smallest cluster sizes remain stable, unlike the earlier calculation of Na inside a defect-free pore (Figure 2) where small clusters are thermodynamically unstable. This distinction is central to interpreting pore-filling kinetics. In a real closed pore, defects on the pore wall can locally lower the barrier to forming the first stable Na nucleus. After this nucleus forms, subsequent filling no longer depends solely on the strength of individual Na-C binding sites but on the ability of Na atoms to diffuse, Na clusters to coalesce, and Na atoms to wet the pore wall. Motivated by the above understanding, we next develop a simplified two-dimensional hard carbon pore model for Na storage (Figure 4a). In this model, we account for different Na-carbon interactions as pair interactions, parametrized by defect-free C- and defective C-Na interactions (Figure 4a, Na interaction energies in the pore follow the qualitative order $|E_{Na-C}|<|E_{Na-Na}|<|E_{Na+D}|$, Na-defective C binding is the strongest). The model uses a grand canonical ensemble (μVT) to simulate the pore-filling process. The pore is modeled as a cylindrical opening inside hard carbon, enclosed within a defective carbon wall. The pore is allowed to exchange Na atoms adjacent to the pore walls with the external Na counterelectrode, which is kept at a constant chemical potential set by the applied voltage. With the interaction energies obtained from the earlier DFT calculations, in this Monte Carlo model Na atoms may diffuse inside, enter, or leave the pore depending on the local pore environment – other Na atoms inside the pore, the pore wall, and defects on the pore wall. We use this model to explore how pore size, voltage, and defect interactions affect the pore-filling voltage and relative kinetics of Na filling. To accelerate diffusion and facilitate configurational sampling, the kinetic simulations use elevated temperatures (2200 K unless otherwise stated), a common approach in theoretical simulations.[33,43,44] This approach provides a common basis for comparing how pore size, defects, and applied voltage influence relative filling kinetics, with numerical results interpreted within the model assumptions described in the Supplementary Information.

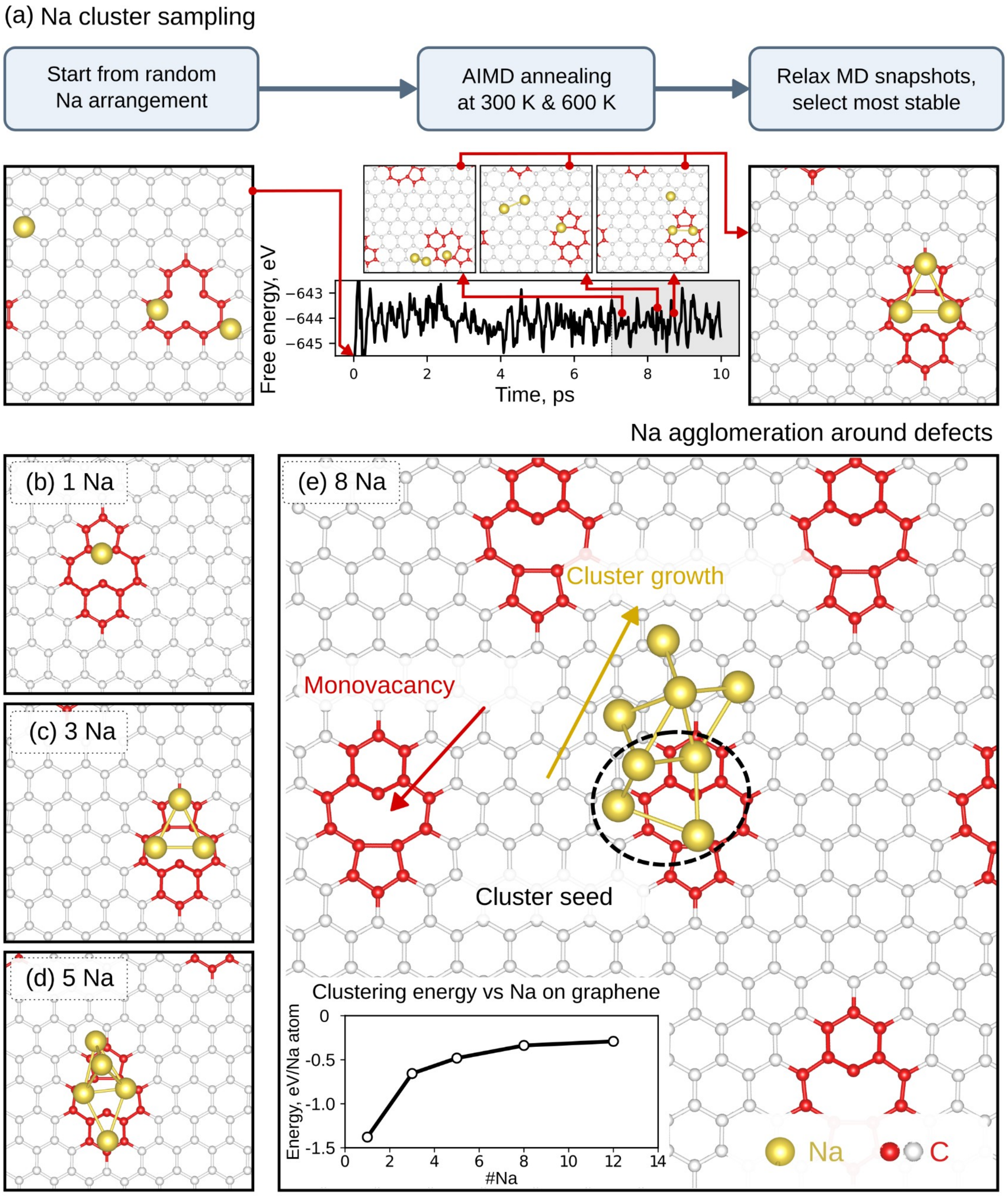

**Figure 3.** Na cluster growth around pre-existing defects on a defective graphene sheet. (a) The workflow to obtain the most stable Na clusters. An individual Na atom strongly binds to a monovacancy (b), forming a stable Na cluster as more Na atoms are adsorbed onto graphene (c-d). The Na cluster serves as a seed for subsequent Na adsorption (e). The inset in (e) shows the energy difference between Na in a cluster relative to individual Na atoms deposited onto graphene away from defects.

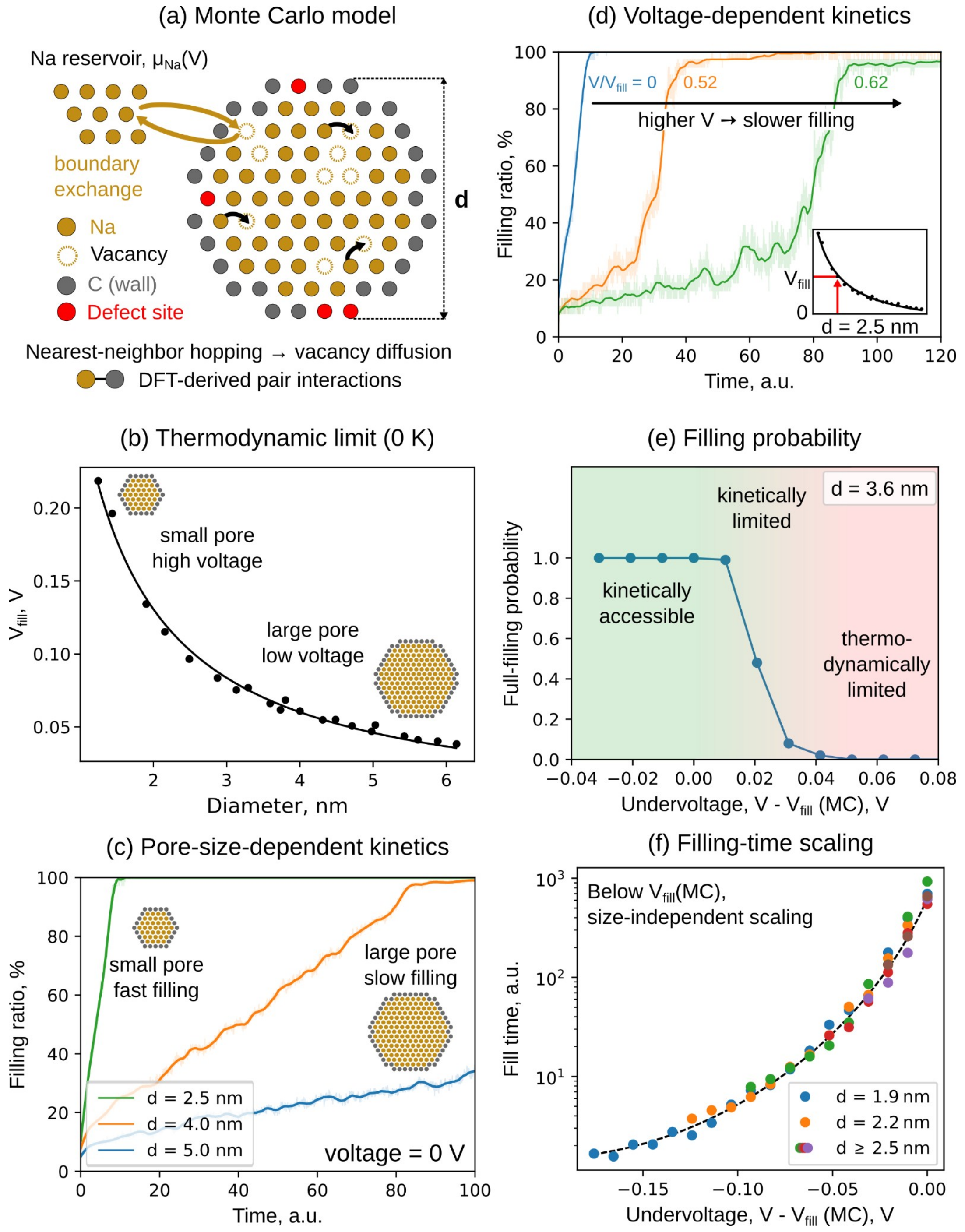

**Figure 4.** Monte Carlo (MC) model of pore filling. (a) Schematic of the MC setup. (b) Voltage of full pore filling for different pore diameters at the thermodynamic limit at 0 K, without MC sampling. (c) Representative MC pore-filling trajectories for different pore sizes. (d) Sample pore-filling trajectory for different voltages. (e) Probability of full pore filling within MC time window as a function of voltage. (f) Scaling of pore-filling time with undervoltage for pores with different diameters. In (c) and (d), thin opaque lines represent median filling ratio across the pore ensemble, and thick solid lines represent smoothed curves over the medians. The data in (b)-(f) is for defect-free pores.

By design, the model predicts the thermodynamic limit of voltage dependence of a fully filled pore on pore radius (Figure 4b). The exact voltage values deviate from those observed in the DFT calculations (Figure 2c), but the model captures the main physics of the Na pore filling – larger pores are filled at lower voltages. The voltage for complete pore filling scales approximately as 0.28/d (where d is the pore diameter in nm; Figure 4b, the inverse pore size scaling is consistent with a previously reported experimental fit for pore-filling energies[45]). The thermodynamic limit calculated from pair interactions (or from DFT) does not represent the entire pore-filling pathway. During the actual filling process, the system must pass through partially filled configurations, which can differ significantly from the fully filled state and may be thermodynamically unfavorable (as shown by the earlier convex hull results in Figure 2b,c). Because of this, positive filling voltage does not guarantee that the pore will fill under finite-time conditions, which becomes particularly clear when the Monte Carlo calculations are performed at a finite time scale. The pore-filling kinetics is strongly pore-size-dependent, with small pores filling significantly faster than the larger ones (Figure 4c, Figure S4). For instance, for 2.5 and 5.0 nm diameter pores, the ratio of median pore-filling times is 1:31. For smaller pores, confinement helps stabilize a smaller intermediate Na cluster nucleus (Figure 2b), so filling can still proceed over a relatively short time. For larger pores, however, larger Na clusters are stable, and the system can remain trapped at low filling ratios for a long time. Because of this, the thermodynamic limit of the voltage may not always be reachable due to a kinetically limited process.

The results presented in Figure 4c are for operations at 0 V – a fundamental condition in theoretical solid-state physics that does not reflect typical battery operating conditions. Experimentally, discharging to 0 V is usually avoided in practical battery operation because of potential Na plating. The developed Monte Carlo model, however, accounts for the charge-discharge process at finite voltage by incorporating the electrochemical Na potential (i.e., the chemical potential of Na outside the pore is offset by the applied voltage: $\mu(\text{Na, electrode}) = \mu(\text{Na, metal}) - eV$), thereby capturing the voltage effect on the pore-filling process. Using a 2.5 nm pore as an example (Figure 4d), we find that the voltage has a substantial effect on the pore-filling kinetics – as the voltage increases from 0 to 0.05 to 0.06 V, the normalized time to fill up the pore increases from 1 to 5.3 to 13.3, respectively. The same trend is observed for other pore sizes, as shown in Figure S5. These results demonstrate that the applied voltage affects not only the thermodynamic limit for pore filling but also the timescale over which filling occurs. The applied voltage contribution can be classified into three regions: (i) kinetically accessible, (ii) kinetically limited, and (iii) thermodynamically limited. As shown in Figure 4e and S5c, at low voltages, pores are kinetically accessible, reaching a full-filling probability of 100% within the simulated time window. At higher voltages, the filling probability drops below 100%, with some pores requiring filling times above 106 MC steps – a kinetically limited regime. The fully filled pore remains thermodynamically favorable, but it is not kinetically accessible within the simulation time. This behavior differs from that of traditional electrode materials where the equilibrium reaction voltage usually simply determines which thermodynamically favorable reactions are activated. Specifically, in the case of pore filling, since the endpoint is only reachable by progressing through a range of unstable pore-filling stages (each of which tends toward decomposition, Figure 2b,c), the applied voltage directly affects the reference points, thereby introducing a kinetic factor into the reaction. Figure 4f further highlights that voltage is the main driving force governing pore-filling kinetics. The filling time is determined by undervoltage – the difference between the applied voltage and the maximum filling voltage within the kinetically accessible region (Figure 4e), irrespective of the pore sizes (within the investigated size range). The difference between pore sizes manifests itself by narrowing the kinetically accessible region between 0 V and the size-dependent maximum voltage. Thus, the applied voltage controls not only the final equilibrium filling state but also whether the system can overcome the nucleation-like bottleneck required to reach that state. These aspects are particularly important for battery operation, as the charge/discharge process typically occurs above 0 V to prevent Na plating and dendrite formation.

While the above cases illustrate pore filling in idealized pore structures, real hard carbons contain a high density of point defects and diverse local structural motifs, which can substantially modify the pore-filling process.[3] First, thermodynamically, increasing the defect concentration raises the filling voltage across the investigated pore-size range because strong Na-defect interactions stabilize Na within the pore (Figure 5a, S6, S7) and accelerate the filling (Figure S8, S9). Thus, defects allow filling larger pores that cannot otherwise be filled (Figure 5b and Figure S10), at least within the considered length of the simulation. In kinetic terms, the defect effect is fundamentally nonmonotonic (Figure 5c and Figure S11). On the one hand, defects accelerate the initial stage of pore filling, which is indeed one of the main limiting stages across the pore-size range. On the other hand, at the late stages of pore filling, when Na atoms wet the pore walls, defects slow down the filling. At the late stages of pore filling, especially in the larger pores, the pore-filling process is dominated by vacancy diffusion, which requires vacancy formation at the pore walls. Since the defect sites bind Na atoms more strongly than the defect-free surface, the presence of defects can significantly reduce the active surface on which vacancy diffusion can be observed, thus increasing the pore-filling time. It should, however, be noted that the exact pore-filling mechanism here may be affected by multiple factors, including the type of defects, interaction energies, pore size, and temperature (see Supplementary Information). However, the same qualitative size dependence is retained at different temperatures (Figure S8e) and the general trend demonstrated in Figure 5c,d holds for all the pore sizes and temperatures we examined (Figure S12) – there is an optimal concentration of point defects that minimizes the pore-filling time. The results thus demonstrate that the pore-filling process requires not only tuning pore size but also addressing defect chemistry. Importantly, the larger pores here do not necessarily provide a higher Na storage capacity because the Na diffusion is kinetically limited.

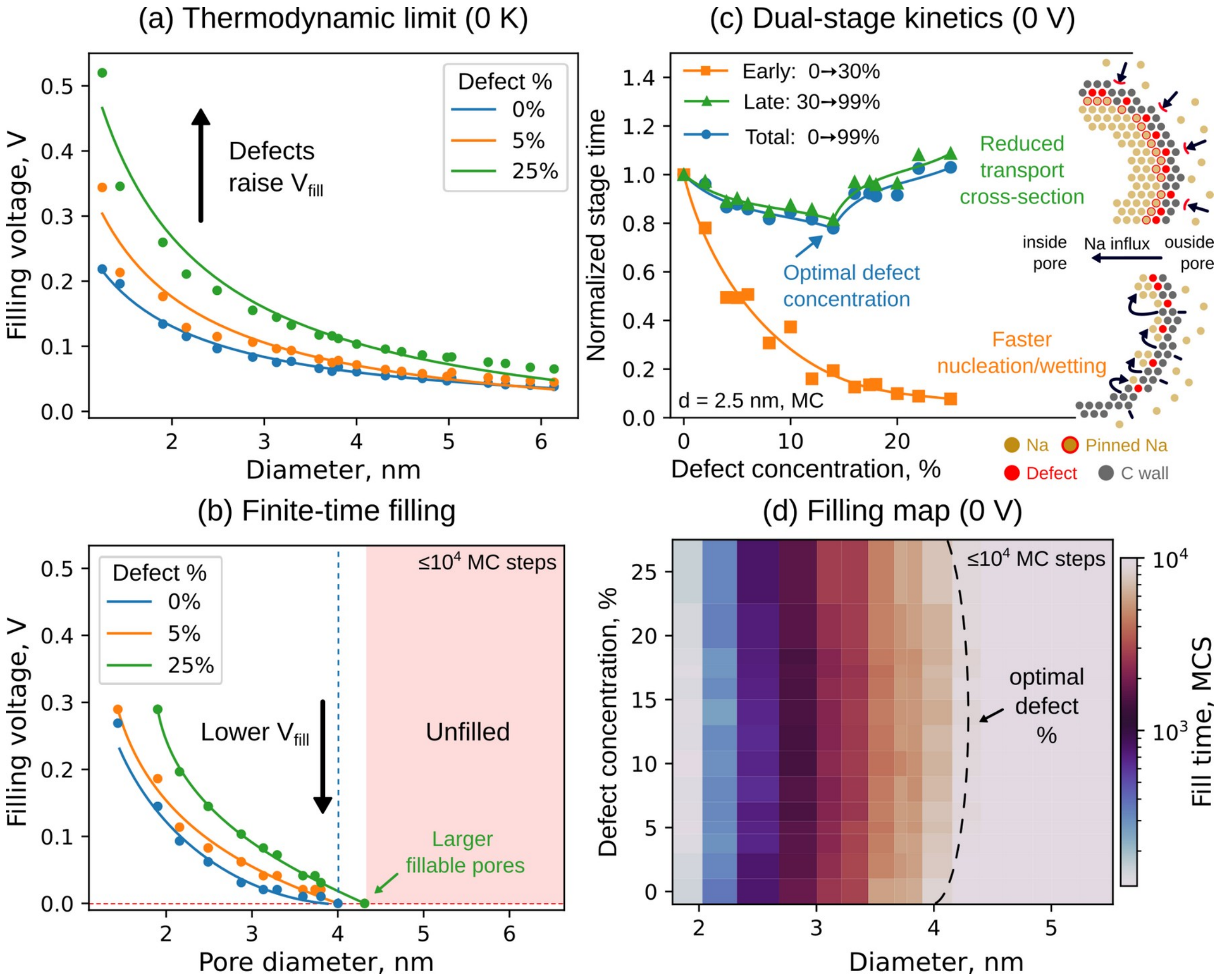

**Figure 5.** Influence of defects on pore filling. (a) Filling voltage at different defect concentrations and pore sizes calculated without the Monte Carlo (MC) model, from a convex hull at 0 K. (b) Filling voltage calculated accounting for filling kinetics, using the Monte Carlo model. The filling voltage is the maximum voltage at which the pore can reach 99% filling within 104 Monte Carlo steps. (c) Effect of defects on early (from 0% to 30%) and late (from 30% to 99%) stages of pore filling according to a Monte Carlo simulation of a pore with a diameter of 2.5 nm. The plots are filling times at different defect concentrations normalized by the corresponding filling time at 0% defects. The schematic illustrates the possible role of defects on pore wall wetting and Na diffusion. (d) Kinetically-limited pore-filling map of pores with different sizes at different defect concentrations. The map shows median filling times for all pores that can be filled within 104 Monte Carlo steps (MCS). Due to the discrete nature of the simulated pores, the actual concentrations vary slightly with pore size.

The mechanistic picture of the pore-filling process is characterized by a sequence of Na cluster nucleation, coalescence, growth, and Na bulk diffusion, as illustrated in Figure S1 and Supplementary Videos S1-S4. At small pore sizes, Na clusters on the pore wall can nucleate relatively quickly, leading to the formation of a single small, stable metallic cluster that grows by accumulating Na atoms. However, when the pore size increases, in the absence of defects, the Na cluster needs to be much larger in order to stabilize, consistent with previous DFT results in Figure 2b,c. Such a large cluster requires significantly more time or a higher thermodynamic driving force (undervoltage) to form. In very large defect-free pores, even the lowest realistic applied voltage (0 V) cannot provide sufficient driving force to fill the pore within a realistic time. When defects are present, they serve as nucleation sites for multiple small and stable clusters. Then, pore-filling kinetics becomes controlled by cluster coalescence, and larger pores can be filled faster. However, cluster coalescence tends to leave the pores with Na fully wetting the pore walls and a large cavity inside. After wetting, the pore filling becomes Na-diffusion-controlled. This process has kinetics that is very different from the nucleation/coalescence stage, as the addition of new Na atoms at this stage is primarily governed by the presence of Na vacancies on the pore wall. The defects pin Na near the pore walls, thereby inhibiting vacancies and slowing diffusion.

## 3. Conclusion

Using first-principles calculations and Monte Carlo simulations, we develop a fundamental understanding of the pore-filling process, one of the most important stages in Na storage in hard carbon. Our results demonstrate that, although many people refer to pore filling as a plateau-type reaction, Na storage in pores actually happens at size-dependent voltages, with different pores contributing differently. Specifically, larger pores can accumulate more Na atoms, but their filling is usually activated at much lower voltages. The pore activation thermodynamic voltage scales roughly as 1/d. Defects activate the pore at higher voltages, significantly accelerating Na wetting the pore walls, but they slow Na diffusion due to reduced Na atom flux over the pore walls (Na is pinned by defects on the wall surface). These findings show that total closed-pore volume alone is insufficient to predict usable plateau-like capacity. Indeed, the design of hard carbon materials should focus on maximizing the kinetically accessible closed-pore volume by jointly controlling pore-size distribution, defect density, and defect-binding strength at a given operating voltage. This thus suggests that simply reducing defect concentration or increasing pore size is not the optimal design strategy for maximizing accessible Na storage capacity.

**Acknowledgements**

The work in China was supported by the Startup Fund of Qingyuan Innovation Laboratory (Grant No. 00524009), the Fujian Province High-Level Innovative and Entrepreneurial Talents Program (Grant No. 30225003), the National Key Research and Development Program of China (Grant No. 2024YFE0101600), and Natural Science Foundation of Fujian Province (Grant No. 2026l0007). The work in Poland was supported by the National Centre for Research and Development under the project WPC3/2022/50/KEYTECH/2024. Institutional and infrastructural support for the ENSEMBLE3 Centre of Excellence was provided through the "Centre of ExcelleNce for nanophotonicS, advancEd Materials and novel crystal growth-Based technoLogiEs - ENSEMBLE3" project which is carried out within the 2.1 International Research Agendas programme of the Foundation for Polish Science co-financed by the European Union under the European Funds for Smart Economy 2021-2027 (FENG) and the European Union Horizon 2020 research and innovation program Teaming for Excellence (GA. No. 857543). The publication was created as part of the project of the Minister of Science and Higher Education "Support for the activities of Centers of Excellence established in Poland under the Horizon 2020 program" under contract No. MEiN/2023/DIR/3797. We acknowledge the Polish high-performance computing infrastructure, PLGrid, for granting this project access to computational resources under project number PLG/2026/019890, as well as to the LUMI supercomputer, owned by the EuroHPC Joint Undertaking and hosted by CSC (Finland) and the LUMI consortium, through project number PLL/2025/09/018974.